\documentclass[a4,12pt]{article}
\usepackage{epsfig}

\begin{document}

\title{Decay Constants of Pseudo-scalar Mesons in Bethe-Salpeter Framework with
Generalized Structure of Hadron-Quark Vertex}
\author{Shashank Bhatnagar \thanks{%
email: shashank$\_$bhatnagar@yahoo.com} \\
{\small Department of Physics, Addis Ababa University}\\
{\small P.O.Box 101739, Addis Ababa, Ethiopia $and$ }\\
{\small Abdus Salam International Center for Theoretical Physics (ICTP),
Trieste, Italy \thanks{%
Regular Associate} }\\
Shi-Yuan Li \thanks{%
email: lishy@sdu.edu.cn} \\
{\small Department of Physics, Shandong University}\\
{\small 250100, Jinan, People's Republic of China $and$ }\\
{\small Theory Division, CERN, CH-1211 Geneva 23, Switzerland}}
\maketitle

\begin{abstract}
We employ the framework of Bethe-Salpeter equation under Covariant
Instantaneous Ansatz to study the leptonic decays of pseudo-scalar mesons.
The Dirac structure of hadron-quark vertex function $\Gamma $ is generalized
to include various Dirac covariants besides $\gamma _5 $ from their complete
set. The covariants are incorporated in accordance with a power counting
rule, order by order in powers of the inverse of the meson mass. The decay
constants are calculated with the incorporation of leading order covariants.
Most of the results are dramatically improved.
\end{abstract}


\section{Introduction}

Quantum Chromodynamics (QCD) is regarded as the correct theory for strong
interactions. Investigation of bound states of quarks (and/or gluons) is one
of the most effective methods to study this dynamics among these
constituents. Since the task of calculating hadron structure from QCD itself
is very difficult (as can be seen from various lattice QCD methods), one can
on the other hand rely on specific models to gain some understanding of it
at low energies, and this study can most effectively be accomplished by
applying a particular framework to a diverse range of phenomena. Meson
decays provide an important opportunity for exploring the structure of these
simplest bound states in QCD and for studying the non-perturbative (long
distance) aspect of the strong interactions. Besides vector mesons (see,
e.g., \cite{1} and refs. therein), pseudo-scalar mesons have for a long time
been a major focus of attention to understand the inner structure of hadrons
from non-perturbative QCD. A number of such studies \cite{2,3,4,5,6,7}
dealing with decays of pseudo-scalar mesons at quark level of compositeness
have been carried out recently. In this paper, we study leptonic decays of
pseudo-scalar mesons such as $K$, $D$, $D_{S}$ and $B$, which proceed through
the coupling of the quark-antiquark loop to the axial vector current as
shown in Figure 1.

A realistic description of pseudo-scalar mesons at the quark level of
compositeness would be an important element in our understanding of hadron
dynamics and reaction processes at scales where QCD degrees of freedom are
relevant. Such studies offer a direct probe of hadron structure and help in
revealing some aspects of the underlying quark-gluon dynamics. The
relativistic description for analyzing mesons as composite objects is
provided by the framework of Bethe-Salpeter Equation (BSE) in this paper. We
employ the QCD oriented BSE under Covariant Instantaneous Ansatz (CIA) \cite
{8}. CIA is a Lorentz-covariant generalization of Instantaneous
Approximation. For $q\bar{q}$ system, CIA formulation \cite{8} ensures an
exact interconnection between 3D and 4D forms of the BSE. The 3D form of BSE
serves for making contact with the mass spectra, whereas the 4D form
provides the $Hq\bar{q}$ vertex function $\Gamma (\widehat{q})$ for the
evaluation of various transition amplitudes. A BSE framework under
Instantaneous Approximation formulation similar to the CIA formulation was
also earlier suggested by the Bonn group \cite{9}.

We had earlier employed the framework of BSE under CIA for calculation of
decay constants \cite{8,10} of heavy-light pseudo-scalar mesons and $F_{\pi }$
for $\pi ^{0}\rightarrow 2\gamma $ process. We also evaluated the leptonic
decays of vector mesons, such as $\rho $, $\omega $, $\phi $ \cite{11} in
this framework. However, one of the simplified assumptions in all these
calculations was that the $Hq\bar{q}$ vertex was restricted to have a single
Dirac structure (e.g., $\gamma _{5}$ for pseudo-scalar mesons, $\gamma \cdot
\varepsilon $ for vector mesons, etc.). However, recent studies \cite
{5,12,13} have revealed that various mesons have many different covariant
structures in their wave functions whose inclusion was also found necessary
to obtain quantitatively accurate observables \cite{12} and it was further
noticed that all Dirac covariants do not contribute equally and only some of
them are relevant for calculation of meson mass spectrum and decay
constants. Such a copious Dirac structure of the BS wave function in fact
was already indicated by Llewellyn Smith \cite{14}. Hence it is necessary to
introduce various Dirac structures into the $Hq\bar{q}$ vertex for different
kinds of mesons. In the recent work \cite{1}, we developed a power counting
rule for incorporating various Dirac covariants in the structure of vertex
function, order by order in powers of inverse of meson mass, and calculated
the leptonic decays of equal mass vector mesons such as $\rho $, $\omega $, $%
\phi $, taking into account the leading order covariants since they are
expected to contribute maximum to observables according to our scheme. On
the line of that work, in this paper we first discuss the power counting
rule for choosing various Dirac covariants from their complete set (see,
e.g., \cite{5,12,13,14}) for pseudo-scalar mesons in Section 2. In section 3
we calculate leptonic decay constants of them employing the wave function
developed in section 2. We then conclude with discussions in Section 4.

\section{Structure of generalized vertex function $\Gamma (\widehat{q})$ for
pseudo-scalar mesons in BSE under CIA}

For introducing the variables and for the convenience to discuss the
generalized hadron-quark vertex in BS wave function under CIA, we give the
similar outline for CIA as in Ref. \cite{1,11}. We start with a 4D BSE for
scalar $q\bar{q}$ system with an effective kernel $K$ and 4D wave function $%
\Phi (P,q)$: 
\begin{equation}
i(2\pi )^{4}\Delta _{1}\Delta _{2}\Phi (P,q)=\int {d^{4}}q^{\prime
}K(q,q^{\prime })\Phi (P,q^{\prime }),  \label{eq1}
\end{equation}
where $\Delta _{1,2}$, the inverse propagators of two scalar quarks, are: 
\begin{equation}
\Delta _{1,2}=m_{1,2}^{2}+p_{1,2}^{2}.  \label{eq2}
\end{equation}
Here $m_{1,2}$ are (effective) constituent masses of quarks. The 4-momenta
of the quark and anti-quark, $p_{1,2}$, are related to the internal
4-momentum $q_{\mu }$ and total momentum $P$ of hadron of mass $M$ as 
\begin{equation}
p_{1,2}{}_{\mu }=\hat{m}_{1,2}P_{\mu }\pm q_{\mu },  \label{eq3}
\end{equation}
where $\hat{m}_{1,2}=[1\pm (m_{1}^{2}-m_{2}^{2})/M^{2}]/2$ are the
Wightman-Garding (WG) definitions \cite{10} of masses of individual quarks.

The CIA Ansatz on the BS kernel $K$ in Eq. (\ref{eq1}) is, 
\begin{equation}
K(q,q^{\prime })=K(\hat{q},\hat{q}^{\prime }),  \label{eq4}
\end{equation}
where 
\begin{equation}
\hat{q}_{\mu }=q_{\mu }-\frac{q \cdot P}{P^{2}}P_{\mu }  \label{eq5}
\end{equation}
is observed to be orthogonal to the total 4-momentum $P$ ( i.e., $\hat{q}%
.P=0)$, irrespective of whether the individual quarks are on-shell or
off-shell. A similar form of the BS kernel was also suggested in ref. \cite
{9}. The longitudinal component of $q_{\mu }$, 
\begin{equation}
M\sigma =M\frac{q \cdot P}{P^{2}},  \label{eq6}
\end{equation}
does not appear in the form $K(\hat{q},\hat{q}^{\prime })$ of the kernel.
For reducing Eq. (\ref{eq1}) to the 3D form, one can define a 3D wave
function $\phi (\widehat{q})$ as 
\begin{equation}
\phi (\widehat{q})=\int\limits_{-\infty }^{+\infty }{Md\sigma \Phi (P,q)}.
\label{eq7}
\end{equation}

Following usual steps outlined in \cite{1,8,11}, we get the $Hq\bar{q}$
vertex function $\Gamma (\hat{q})$ under CIA for the case of scalar quarks: 
\begin{equation}
\Delta _{1}\Delta _{2}\Phi (P,q)=\frac{D(\widehat{q})\phi (\widehat{q})}{%
2\pi i}\equiv \Gamma (\widehat{q})\quad ~ \quad \frac{1}{D(\widehat{q})}=%
\frac{1}{2\pi i}\int\limits_{-\infty }^{+\infty }{\frac{Md\sigma }{\Delta
_{1}\Delta _{2}}},  \label{eq8}
\end{equation}
where $D(\widehat{q})$ is a 3D denominator function whose value can be
easily worked out by contour integration by noting the positions of the
poles in the complex $\sigma -$plane \cite{1,8,11}. By this process, an
exact interconnection between 3D wave function $\phi (\widehat{q})$ and the
4D wave function $\Phi (P,q)$ and hence that between 3D and 4D BSE is thus
brought out, where the 3D form serves for making contact with the mass
spectrum of hadrons, whereas the 4D form provides the $Hq\bar{q}$ vertex
function $\Gamma (\hat{q})$ which satisfies a 4D BSE with a natural
off-shell extension over the entire 4D space (due to the positive
definiteness of the quantity $\hat{q}^{2}=q^{2}-\frac{(q\cdot P)^{2}}{P^{2}}$
throughout the entire 4D space) and thus provides a fully Lorentz-covariant
basis for evaluation of various transition amplitudes through various quark
loop diagrams (see Figure 1).

To apply the above simplified discussions to the case of fermionic quarks
constituting a particular meson we proceed in the same manner as \cite{1}:
The scalar propagators $\Delta _{i}^{-1}$ in the above equations are
replaced by the proper fermionic propagators $S_{F}$. Then, on observing the
vertex $\Gamma (\hat{q})$ now is a $4\times 4$ matrix in the spinor space,
we should incorporate its relevant Dirac structures, for which we take
guidance from some recent studies \cite{5,12,13}, as well as
Llewelyn-Smith's classic paper \cite{14}, which have revealed that various
mesons have many different covariant structures in their wave functions
whose inclusion was found necessary to obtain quantitatively accurate
observables. It was also noticed recently \cite{12} that all Dirac
covariants do not contribute equally and only some covariants are considered
to be relevant for calculation of mass spectrum and decay constants. Ref. 
\cite{12} has also calculated masses and decay constants of vector mesons
for various subsets of covariants from their complete set. Towards this end,
we make use of the power counting rule developed in \cite{1} for
incorporating various Dirac covariants in the structure of $Hq\bar{q}$
vertex function for a particular meson, order-by-order in powers of inverse
of the meson mass $M$, so as to systematically choose among various
covariants from their complete set and write wave functions for various
mesons.

As far as a pseudo-scalar meson is concerned, its $Hq\bar{q}$ vertex
function, which has a certain dimensionality of mass, can be expressed as a
linear combination of four Dirac covariants $\Gamma _{i}^{P}(i=0,...,3)$ 
\cite{12,13,14}, each multiplying a Lorentz scalar amplitude $%
F_{i}(q^{2},q\cdot P,P^{2})$. The choice of the Dirac covariants is not
unique as can also be seen from the choice of covariants used in Ref. \cite
{12,13}. For adapting this decomposition to write the structure of vertex
function $\Gamma (\hat{q})$, we re-express the $Hq\bar{q}$ vertex function
by making the amplitudes $F_{i}(q^{2},q\cdot P,P^{2})$ dimensionless,
weighing each Dirac covariant with an appropriate power of $M$. Thus each
term in the expansion of $\Gamma (\hat{q})$ is associated with a certain
power of $M$. In detail, we can express $\Gamma _{P}$ as a polynomial in
various powers of $1/M$: 
\begin{equation}
\Gamma _{P}=\Omega _{P}\frac{1}{2\pi i}N_{P}D(\hat{q})\phi (\hat{q}),
\label{eq10}
\end{equation}
with 
\begin{equation}
\Omega _{P}=\gamma _{5}B_{0}-i\gamma _{5}(\gamma \cdot P)\frac{B_{1}}{M}%
-i\gamma _{5}(\gamma \cdot q)\frac{B_{2}}{M}-\gamma _{5}(\gamma \cdot
P\gamma \cdot q-\gamma \cdot q\gamma \cdot P)\frac{B_{3}}{M^{2}}.
\end{equation}
Here $B_{i}(i=0,...,3)$ are four dimensionless and constant coefficients
(which are taken to be constant on lines of \cite{1}, only to consider the
leading powers of $1/M$) to be determined. Now since we use constituent
quark masses where the quark mass $m$ is approximately half of the hadron
mass $M$, we can use the ansatz 
\begin{equation}
q<<P\sim M  \label{eq11}
\end{equation}
in the rest frame of the hadron (among all the pseudo-scalar mesons, pion
enjoys the special status in view of its unusually small mass ($<\Lambda
_{QCD}$) and its case should be considered separately. See the discussions
in Section 4). Then each of the four terms in Eq. (\ref{eq10}) receives
suppression by different powers of $1/M$. Thus we can arrange these terms as
an expansion in powers of $O(\frac{1}{M})$. We can see in the expansion of $%
\Omega _{P}$ that the structures associated with the coefficients $%
B_{0},B_{1}$ have magnitudes $O(\frac{1}{M^{0}})$ and are of leading order,
while those with $B_{2},B_{3}$ are $O(\frac{1}{M^{1}})$and are
next-to-leading order. This na\"{i}ve power counting rule suggests that the
maximum contribution to the calculation of any pseudo-scalar meson observable
should come from the leading order Dirac structures $\gamma _{5}$ and $%
i\gamma _{5}(\gamma .P)\frac{1}{M}$ associated with the constant
coefficients $B_{0}$ and $B_{1}$, respectively. As a first application of
this to the pseudo-scalar meson sector and on lines similar to \cite{1} for
vector meson case, we take the form of the $Hq\bar{q}$ vertex function
incorporating the leading order terms in expansion (\ref{eq10}) and ignoring 
$O(\frac{1}{M^{1}})$ terms for the moment, i.e., 
\begin{equation}
\Gamma (\hat{q})=[\gamma _{5}B_{0}-i\gamma _{5}\gamma \cdot P\frac{B_{1}}{M}]%
\frac{1}{2\pi i}N_{P}D(\hat{q})\phi (\hat{q})\quad .  \label{eq12}
\end{equation}
As has been stated in \cite{1}, the restriction from charge Parity on the
wave function of the eigenstate should be respected \footnote{%
To get the complete set of the Dirac structure for a certain kind of mesons,
the restriction by the (space) Parity have been employed; and it is easy to
see that the requirements of the space Parity and the charge Parity are the
same for the vertex as well as the full wave function (see also \cite{14}).}.

From the above analysis of the structure of $Hq\bar{q}$ vertex function in
Eq.(\ref{eq12}) we notice that, at leading order, the structure of 3D wave
function $\phi (\hat{q})$ as well as the form of the 3D BSE are left
untouched and have the same form as in our previous works, which justifies
the usage of the same form of the input kernel we used earlier. Now we
briefly mention some features of the BS formulation employed. The structure
of BSE is characterized by a single effective kernel arising out of a
four-fermion Lagrangian in the Nambu-Jonalasino \cite{15,16} sense. The
formalism is fully consistent with Nambu-Jona-Lasino \cite{16} picture of
chiral symmetry breaking but is additionally Lorentz-invariant because of
the unique properties of the quantity $\hat{q}^{2}$, which is positive
definite throughout the entire 4D space. The input kernel $K(q,q^{\prime })$
in BSE is taken as one-gluon-exchange like as regards color ($\frac{1}{2}%
\vec{\lambda}.\frac{1}{2}\vec{\lambda}_{2})$ and spin ($\gamma _{\mu
}^{(1)}\gamma _{\mu }^{(2)})$ dependence. The scalar function $V(q-q^{\prime
})$ is a sum of one-gluon exchange $V_{OGE}$ and a confining term $V_{Conf.}$
\cite{1,11,15}. Thus

\[
K(q,q^{\prime}) = \frac{1}{2}\vec {\lambda }^{(1)}\frac{1}{2}\vec {\lambda
}^{(2)}V_\mu ^{(1)} V_\mu ^{(2)} V(q - q^{\prime}); 
\]

\[
V_\mu ^{(1,2)} = \pm 2m_{1,2} \gamma _\mu ^{(1,2)} ; 
\]

\begin{equation}
V(\hat{q}-\hat{q}^{\prime }) =\frac{4\pi \alpha _{S}(Q^{2})}{(\hat{q}-\hat{%
q}^{\prime })^{2}}+\frac{3}{4}\omega _{q\bar{q}}^{2}\int {d^{3}\vec{r}[r^{2}}%
(1+4a_{0}\hat{m}_{1}\hat{m}_{2}M_{>}^{2}r^{2})^{-\frac{1}{2}}-\frac{C_{0}}{%
\omega _{0}^{2}}]e^{i(\hat{q}-\hat{q}^{\prime })\cdot \vec{r}};  \label{eq13}
\end{equation}

\begin{equation}
\alpha _{S}(Q^{2}) =\frac{12\pi }{33-2f}\left[ {\ln \frac{M_{>}^{2}}{%
\Lambda ^{2}}}\right] ^{-1};~~M_{>}=Max(M,m_{1}+m_{2}).
\end{equation}

The ansatz employed for the spring constant $\omega _{q\overline{q}}^{2}$ in
the above equation is [1,11,15]

\begin{equation}
\omega _{q\overline{q}}^{2}=4\widehat{m}_{1}\widehat{m}_{2}M_{>}\omega
_{0}^{2}\alpha _{S}(Q^{2}).  \label{eq14}
\end{equation}

Here the proportionality of $\omega _{q\overline{q}}^{2}$ on $\alpha
_{S}(Q^{2})$ is needed to provide a more direct QCD motivation to
confinement. This assumption further facilitates a flavour variation in $%
\omega _{q\overline{q}}^{2}$. And $\omega _{0}^{2}$ in Eq.(13) and Eq.(14)
is postulated as a universal spring constant which is common to all flavours.

In the expression for $V(\hat{q}-\hat{q}^{\prime })$, as far as the
integrand of confining term $V_{Conf.}$ is concerned, the constant term $%
C_{0}/\omega _{0}^{2}$ is designed to take into account the correct zero
point energies, while $a_{0}$ term ($a_{0}<<1)$ simulates an effect of an
almost linear confinement for heavy quark sectors (large $m_{1},m_{2}$ ),
retaining the harmonic form for light quark sectors (small $m_{1},m_{2})$ 
\cite{1,15}, as is believed to be true for QCD. Hence the term ${r^{2}}%
(1+4a_{0}\hat{m}_{1}\hat{m}_{2}M_{>}^{2}r^{2})^{-\frac{1}{2}}$ in the above
expression is responsible for effecting a smooth transition from harmonic ($q%
\overline{q}$) to linear ($Q\overline{Q}$) confinement. The values of basic
input parameters of the model are $a_{0}=.028,~C_{0}=.29,~\omega
_{0}=.158 ~(GeV)$ 
   and $\Lambda =.20GeV$ [1,10,11,15] which have been calibrated to fit the $q%
\overline{q}$ hadron mass spectrum obtained by solving the 3D BSE [15].

Now comes to the problem of the 3D BS wave function. The ground state wave
function $\phi (\widehat{q})$ satisfies the 3D BSE \cite{1} on the surface
P.q = 0, which is appropriate for making contact with O(3)-like mass
spectrum \cite{15}. Its fuller structure (described in Ref. \cite{15}) is
reducible to that of a 3D harmonic oscillator with coefficients dependent on
the hadron mass M and the total quantum number N. The ground state wave
function $\phi (\hat{q})$deducible from this equation thus has a Gaussian
structure \cite{1,8,11} and is expressible as: 
\begin{equation}
\phi (\widehat{q})\approx e^{-\widehat{q}^{2}/2\beta ^{2}}.  \label{15}
\end{equation}

In the structure of $\phi (\hat{q})$ in Eq. (\ref{15}), the parameter $%
\beta $ is the inverse range parameter which incorporates the content of BS
dynamics and is dependent on the input kernel $K(q,q^{\prime })$. The
structure of $\beta $ is given in Section 3.

\section{Decays constants $f_{P}$ of Pseudo-scalar Mesons}

Decay constants $f_{P}$ can be evaluated through the loop diagram shown in
Figure 1 which gives the coupling of the two-quark loop to the axial vector
current and can be evaluated as: 
\begin{equation}
f_{P}P_{\mu }=<0|\bar{Q}i\gamma _{\mu }\gamma _{5}Q|P(P)>,  \label{eq16}
\end{equation}
which can in turn be expressed as a loop integral: 
\begin{equation}
f_{P}P_{\mu }=\sqrt{3}\int {d^{4}}qTr[\Psi _{P}(P,q)i\gamma _{\mu }\gamma
_{5}]\quad .  \label{eq17}
\end{equation}

Bethe-Salpeter wave function $\Psi (P,q)$ for a P-meson is expressed as 
\begin{eqnarray}
\Psi (P,q)=S_{F}(p_{1})\Gamma (\hat{q})S_{F}(-p_{2}), &&with~~~~  \nonumber
\\
S_{F}(p_{1})=-i\frac{(m_{1}-i\gamma \cdot p_{1})}{\Delta _{1}};
&&S_{F}(-p_{2})=-i\frac{(m_{2}+i\gamma \cdot p_{2})}{\Delta _{2}}.
\label{eq18}
\end{eqnarray}

In the following calculation, we only take the leading order terms in the
structure of hadron-quark vertex function $\Gamma (\hat{q})$ as in Eq. (\ref
{eq12}). $S_{F}$ are the fermionic propagators for the two constituent
quarks of the hadron and the non-perturbative aspects enter through the $%
\Gamma (\widehat{q})$. Using $\Psi (P,q)$ from Eq.(\ref{eq18}) and the
structure of $Hq\bar{q}$ vertex $\Gamma (\hat{q})$ from Eq.(\ref{eq12}),
evaluating trace over $\gamma $-matrices and multiplying both sides of Eq.(%
\ref{eq17}) by $P_{\mu }/(-M^{2})$, we can express $f_{P}$ as:

\[
f_P = \sqrt 3 N_P \int {d^3} \hat {q}D(\hat {q})\phi (\hat {q})I \quad , 
\]

\begin{eqnarray}
I &=&\int\limits_{-\infty }^{+\infty }{\frac{Md\sigma }{2\pi i\Delta
_{1}\Delta _{2}}}\{2B_{0}[m_{12}(1-\frac{\delta m^{2}}{M^{2}})+2\delta
m\sigma ]  \label{eq19} \\
&+&\frac{B_{1}}{M^{3}}[(-M^{4}-4m_{1}m_{2}M^{2}+m_{12}^{2}\delta
m^{2})+4M^{2}\hat{q}^{2}-4M^{2}m_{12}\delta m\sigma +4M^{4}\sigma ^{2}]\}, 
\nonumber
\end{eqnarray}
\noindent where according to Eq.(\ref{eq3}) and Eq.(\ref{eq5}), we had
expressed scalar products $p_{1}\cdot p_{2}$, $p_{1}\cdot P$and $p_{2}\cdot
P $ as 
\begin{equation}
p_{1}.p_{2}=-M^{2}(\hat{m}_{1}+\sigma )(\hat{m}_{2}-\sigma )-\hat{q}^{2},
\label{eq20}
\end{equation}

\[
\begin{array}{l}
p_{1}\cdot P=-M^{2}(\hat{m}_{1}+\sigma ) \\ 
p_{2}\cdot P=-M^{2}(\hat{m}_{2}-\sigma ),
\end{array}
\]
\noindent with $m_{12}=m_{1}+m_{2}$ and $\delta m=m_{2}-m_{1}$. Here we have
employed unequal mass kinematics when the hadron constituents have different
masses. We see that on the right hand side of the expression for $f_{P}$,
each of the expressions multiplying the constant parameters $B_{0}$ and $%
B_{1}$ consist of two parts, of which only the second part explicitly
involves the off-shell parameter $\sigma $ (that is terms involving $\sigma $
multiplying $B_{0}$, and terms involving both $\sigma $ and $\sigma ^{2}$
multiplying $B_{1})$). It is seen that the off-shell contribution which
vanishes for $m_{1}=m_{2}$ in case of $f_{P}$ calculation in CIA \cite{10}
using only the covariant $\gamma _{5}$, now no longer vanishes for $%
m_{1}=m_{2}$ in the above calculation, due to the term $4M^{4}\sigma ^{2}$
multiplying $B_{1}$ when another leading order covariant $i\gamma _{5}\gamma
\cdot P/M$ is also incorporated in $Hq\bar{q}$ vertex function. This
possibly implies that the off-shell part of $f_{P}$ does not arise from
unequal mass kinematics alone, which is in complete contrast to the earlier
CIA calculation of $f_{P}$ \cite{10} employing only $\gamma _{5}$. This
serves as a clear pointer to the fact that in this BS-CIA model, the Dirac
covariants other than $\gamma _{5}$ might also be important for the study of
processes involving large $q^{2}$ (off-shell), which is also suggested in 
\cite{17}.

Carrying out integration over $d\sigma $ by noting the pole positions in the
complex $\sigma $-plane 
\begin{equation}
\begin{array}{l}
\Delta _{1}=0\Rightarrow \sigma _{1}^{\pm }=\pm \frac{\omega _{1}}{M}-\hat{m}%
_{1}\mp i\varepsilon ,\omega _{1}^{2}=m_{1}^{2}+\hat{q}^{2} \\ 
\Delta _{2}=0\Rightarrow \sigma _{2}^{\mp }=\mp \frac{\omega _{2}}{M}+\hat{m}%
_{2}\pm i\varepsilon ,\omega _{2}^{2}=m_{2}^{2}+\hat{q}^{2},
\end{array}
\label{eq21}
\end{equation}
\noindent we can express $f_{P}$ as 
\begin{eqnarray}
f_{P} &=&\sqrt{3}N_{P}\int d^{3}\hat{q}D(\hat{q})\phi (\hat{q})\Big (%
2B_{0}[m_{12}(1-\frac{\delta m^{2}}{M^{2}})\frac{1}{D(\hat{q})}+2\delta
mR_{1}]  \nonumber \\
&+&\frac{B_{1}}{M^{3}}[(-M^{4}-4m_{1}m_{2}M^{2}+m_{12}^{2}\delta m^{2})\frac{%
1}{D(\hat{q})}  \nonumber \\
&+&\frac{4\hat{q}^{4}}{MD(\hat{q})}(-4M^{2}m_{12}\delta m.R_{1}+4M^{4}R_{2})]%
\Big ),  \label{eq22}
\end{eqnarray}
where the relationship between the functions $D_{0}(\hat{q})$ and $D(\hat{q}%
) $ (see Ref. \cite{11} for details) is $D(\hat{q})=\frac{D_{0}(\hat{q})}{(%
\frac{1}{2\omega _{1}}+\frac{1}{2\omega _{2}})};\quad D_{0}(\hat{q})=(\omega
_{1}+\omega _{2})^{2}-M^{2}$. The results of $\sigma $ contour integration 
is: 
\begin{equation}
R_{1}=\int\limits_{-\infty }^{+\infty }{\frac{Md\sigma }{2\pi i\Delta
_{1}\Delta _{2}}}\sigma =\frac{M^{2}(-\omega _{1}+\omega
_{2})+(m_{1}^{2}-m_{2}^{2})(\omega _{1}+\omega _{2})}{4M^{2}\omega
_{1}\omega _{2}(M^{2}-(\omega _{1}+\omega _{2})^{2})}\quad ;  \label{eq23}
\end{equation}

\begin{eqnarray}
R_{2} &=&\int\limits_{-\infty }^{+\infty }{\frac{Md\sigma }{2\pi i\Delta
_{1}\Delta _{2}}}\sigma ^{2}  \nonumber \\
&=&\frac{(-M^{4}-m_{12}^{2}\delta m^{2}+4M^{2}\omega _{1}\omega _{2})(\omega
_{1}+\omega _{2})+2M^{2}m_{12}\delta m(\omega _{2}-\omega _{1})}{%
8M^{4}\omega _{1}\omega _{2}(M^{2}-(\omega _{1}+\omega _{2})^{2})}.
\end{eqnarray}

The structure of the parameter $\beta $ in $\phi (\widehat{q})$ appearing in
Eq.(\ref{eq22}) is taken from Ref. \cite{1,10,11,15,18}. (for details see
Ref. \cite{1}). Thus we take 
\begin{equation}
\beta ^{2}=(2\hat{m}_{1}\hat{m}_{2}M\omega _{q\bar{q}}^{2}/\gamma
^{2})^{1/2};\gamma ^{2}=1-\frac{2\omega _{q\bar{q}}^{2}C_{0}}{M_{>}\omega
_{0}^{2}}.  \label{eq24}
\end{equation}
Here $\omega _{q\bar{q}}^{2}$ is expressed as in Eq. (\ref{eq14}). To
calculate the normalization factor$N_{P}$, we use the current conservation
condition \cite{8}, 
\begin{equation}
2iP\mu =(2\pi )^{4}\int d^{4}qTr[\overline{\psi }(P,q)(\frac{\partial }{%
\partial P_{\mu }}S_{F}^{-1}(p_{1}))\psi
(P,q)S_{F}^{-1}(-p_{2})]+(1\Leftrightarrow 2),  \label{currcon}
\end{equation}
where the momentum of constituent quarks can be expressed as in Eq.(\ref{eq3}%
). Taking the derivatives of inverse of propagators of constituent quarks
with respect to the total 4-momentum $P_{\mu }$, evaluating trace over the $%
\gamma $-matrices and following usual steps outlined in Ref.\cite{1,18},
then carrying out the $d\sigma $ integral by noting the pole positions in
complex $\sigma $-plane, we can express the normalizer as, 
\begin{eqnarray}
N_{P}^{-2} &=&-(2\pi )^{2}i\int {d^{3}}\widehat{q}D^{2}(\widehat{q})\phi
^{2}(\widehat{q})S\quad ;~~with~  \nonumber \\
S &=&\frac{4B_{0}B_{1}}{M^{3}}[(-\delta
m^{3}m_{12}^{2}-M^{4}m_{12}+2M^{2}m_{12}^{2}\delta m)I_{1}  \nonumber \\
&+&(2M^{4}m_{12}+2M^{4}\delta m+4M^{2}m_{12}\delta m^{2})I_{2}-(4M^{4}\delta
m)I_{3}]  \nonumber \\
&-&\frac{B_{1}^{2}}{M^{4}}[D_{1}E_{1}I_{1}+(2M^{2}E_{1}-4M^{2}\delta
mm_{12}D_{1})I_{2}  \nonumber \\
&+&(-8M^{4}\delta mm_{12}+4M^{4}D_{1})I_{3}  \nonumber \\
&+&(8M^{6})I_{4}-\frac{B_{0}^{2}}{M^{4}}[D_{1}E_{2}I_{1}+(2M^{2}E_{2}-4M^{2}%
\delta mm_{12}D_{1})I_{2}  \nonumber \\
&+&(-8M^{4}\delta mm_{12}+4M^{4}D_{1})I_{3}+(8M^{6})I_{4}]  \nonumber \\
&+&\frac{4B_{0}B_{1}}{M^{3}}[-2M^{2}m_{2}\frac{1}{D(\hat{q})}]  \nonumber \\
&-&\frac{B_{1}^{2}}{M^{4}}[4M^{4}R_{1}+(-2M^{2}m_{12}\delta m+2M^{4})\frac{1%
}{D(\hat{q})}]  \nonumber \\
&-&\frac{B_{0}^{2}}{M^{4}}[-2(M^{2}m_{12}\delta m+M^{4})\frac{1}{D(\hat{q})}%
+4M^{4}R_{1}]  \nonumber \\
&+&\hat{q}^{2}\Big ({\frac{-B_{1}^{2}}{M^{4}}[4M^{2}D_{1}I_{1}+8M^{4}I_{2}]-%
\frac{B_{0}^{2}}{M^{4}}[-4M^{2}D_{1}I_{1}-8M^{4}I_{2}]}\Big ).  \label{28}
\end{eqnarray}

In the above, the quantities $D_{1},E_{1},E_{2}$ are: \bigskip $%
D_{1}=-m_{12}\delta m+M^{2}$; $E_{1}=\delta
m^{2}m_{12}^{2}-4M^{2}m_{1}m_{2}-M^{4}$; $E_{2}=E_{1}-3M^{2}m_{1}m_{2}$,
while the integrals $I_{1},I_{2},I_{3}$ and $I_{4}$ over the off-shell
parameter $d\sigma $ are:

\begin{equation}
I_{1}=\int\limits_{-\infty }^{+\infty }{\frac{Md\sigma }{\Delta
_{1}^{2}\Delta _{2}}=2\pi i\left[ {\frac{2\omega _{1}^{3}-M^{2}\omega
_{2}+5\omega _{1}^{2}\omega _{2}+4\omega _{1}\omega _{2}^{2}+\omega _{2}^{3}%
}{4\omega _{1}^{3}\omega _{2}(M^{2}-(\omega _{1}+\omega _{2})^{2})^{2}}}%
\right] }
\end{equation}
\begin{eqnarray}
I_{2} &=&\int\limits_{-\infty }^{+\infty }\frac{Md\sigma }{\Delta
_{1}^{2}\Delta _{2}}\sigma  \nonumber \\
&=&2\pi i\Big (\frac{-M^{4}\omega _{2}+(m_{1}^{2}-m_{2}^{2})(\omega
_{1}+\omega _{2})^{2}(2\omega _{1}+\omega _{2})}{8M^{2}\omega _{1}^{3}\omega
_{2}(M^{2}-(\omega _{1}+\omega _{2})^{2})^{2}}  \nonumber \\
&+&\frac{M^{2}[6\omega _{1}^{3}+9\omega _{1}^{2}\omega _{2}+4\omega
_{1}\omega _{2}^{2}+\omega _{2}(-m_{1}^{2}+m_{2}^{2}+\omega _{2}^{2})]}{%
8M^{2}\omega _{1}^{3}\omega _{2}(M^{2}-(\omega _{1}+\omega _{2})^{2})^{2}}%
\Big );
\end{eqnarray}
\begin{eqnarray}
I_{3} &=&\int\limits_{-\infty }^{+\infty }\frac{Md\sigma }{\Delta
_{1}^{2}\Delta _{2}}\sigma ^{2}  \nonumber \\
&=&2\pi i\frac{1}{16M^{4}}\Big(\frac{2(M^{2}-m_{1}^{2}+m_{2}^{2}+2M\omega
_{2})^{2}}{\omega _{2}(M-\omega _{1}+\omega _{2})^{2}(M+\omega _{1}+\omega
_{2})^{2}}  \nonumber \\
&+&\frac{2M(M-m_{1})(M^{2}+m_{1}^{2}-m_{2}^{2}-2M\omega _{1})^{2}}{\omega
_{1}^{2}((M-\omega _{1})^{2}-\omega _{2}^{2})^{2}}  \nonumber \\
&-&\frac{4M^{2}(M^{2}+m_{1}^{2}-m_{2}^{2}-2M\omega _{1})}{\omega
_{1}^{2}((M-\omega _{1})^{2}-\omega _{2}^{2})}  \nonumber \\
&-&\frac{M(M^{2}+m_{1}^{2}-m_{2}^{2}-2M\omega _{1})^{2}}{\omega
_{1}^{3}((M-\omega _{1})^{2}-\omega _{2}^{2})}\Big );
\end{eqnarray}

\begin{eqnarray}
I_{4} &=&\int\limits_{-\infty }^{+\infty }{\frac{Md\sigma }{\Delta
_{1}^{2}\Delta _{2}}}\sigma ^{3}  \nonumber \\
&=&2\pi i.\Big (\frac{2(M^{2}-m_{1}^{2}+m_{2}^{2}+2M\omega _{2})^{3}}{%
16M^{6}\omega _{2}(M^{2}-\omega _{1}^{2}+2M\omega _{2}+\omega _{2}^{2})^{2}}
\nonumber \\
&+&\frac{(M^{2}+m_{1}^{2}-m_{2}^{2}-2M\omega
_{1})^{2}(M^{4}+M^{2}(m_{1}^{2}-m_{2}^{2}-\omega _{1}^{2}-\omega _{2}^{2})}{%
16M^{6}\omega _{1}^{3}(M^{2}-2M\omega _{1}+\omega _{1}^{2}-\omega
_{2}^{2})^{2}}  \nonumber \\
&+&\frac{(m_{1}^{2}-m_{2}^{2})(3\omega _{1}^{2}-\omega _{2}^{2})-4M\omega
_{1}(m_{1}^{2}-m_{2}^{2}+\omega _{2}^{2}))}{16M^{6}\omega
_{1}^{3}(M^{2}-2M\omega _{1}+\omega _{1}^{2}-\omega _{2}^{2})^{2}}\Big ).
\end{eqnarray}

We have thus evaluated the general expressions for $f_{P}$ (Eq.(\ref{eq22}))
and $N_{P}$ (Eq.(\ref{28})) in the framework of BSE under CIA, with Dirac
structure $(i\gamma _{5})(\gamma \cdot P)/M$ introduced in the $Hq\bar{q}$
vertex function as the leading order structure as well as $\gamma _{5}$,
according to our power counting rule. We see that so far the results are
independent of any model for $\phi (\widehat{q})$. However, for calculating
the numerical values of these decay constants one needs to know the constant
coefficients $B_{0}$ and $B_{1}$ which are associated with the Dirac
structures $\gamma _{5}$ and $(i\gamma _{5})(\gamma \cdot P)/M$,
respectively. The relative value, $\frac{B_{1}}{B_{0}}$ is a free parameter
without any further knowledge of the meson structure in the framework
discussed above. As a first step, we vary this parameter to see the effect
of introducing the Dirac covariant $i\gamma _{5}\gamma \cdot P/M$. We see
that at $B_{1}/B_{0}=.163$, $f_{K}=159.8MeV$, which is the experimental
value of this quantity. For making comparison with results of other models
and data, we use this value of the ratio $B_{1}/B_{0}$. We also vary $%
B_{1}/B_{0}$ in the range $.14-.17$ to demonstrate the dependence on this
parameter. The results are given in Table I along with those of other models
and experimental data. It is seen that most of the numerical values of these
decay constants in BSE under CIA improve when $\gamma _{5}(\gamma .P)/M$ is
introduced in the vertex function in comparison to the values calculated
with only $\gamma _{5}$. Further discussions are in Sec. 4.

\section{Discussions}

In this paper we have first written $Hq\bar{q}$ vertex function for a
pseudo-scalar meson employing the power counting rule \cite{1} for the
incorporation of various Dirac covariants from their complete set. They are
incorporated order-by-order in powers of inverse of the meson mass. We then
calculate $f_{P}$ for unequal mass pseudo-scalar mesons ($K$, $D$, $D_{S}$, $%
B $) in the framework of Bethe-Salpeter Equation under CIA, using the
hadron-quark vertex function $\Gamma (\widehat{q})$ in Eq. (\ref{eq12}) with
the incorporation of the leading order covariants. Unequal mass kinematics
is employed in this calculation. It is seen that the values of Decay
constants can improve considerably when the Dirac structure $i\gamma
_{5}(\gamma \cdot P)/M$ is introduced in the vertex function with tuned
parameter $\frac{B_{1}}{B_{0}}$, and come closer to the results of some
recent calculations \cite{12, 13,19,20} as well as agree with experimental
results \cite{22} within error though $f_{B}$ is a bit lower (but still
within 1-$\sigma $, for details see Table 1).

In a recent work \cite{12}, the Leptonic decay constants $f_{P}$ have been
calculated for light pseudo-scalar mesons within a ladder-rainbow truncation
of coupled Dyson-Schwinger and Bethe-Salpeter Equations using $Hq\bar{q}$
vertex function to be a linear combination of four dimensionless orthogonal
Dirac covariants where each covariant multiplies a scalar amplitude $%
F_{i}(q^{2},q\cdot P,P^{2})$ for three different parameter sets for
effective interactions. The decay constants for pseudo-scalar meson $K$
calculated in that model \cite{5,12} for an intermediate value of one of the
parameter sets: $\omega =0.4GeV,D=0.93GeV^{2}$ is $f_{K}=155MeV.$ In our
model with leading covariant$\gamma _{5}$ alone we obtain $f_{K}=153.5MeV$,
which is quite close to this figure. However, here we include the other
leading order Dirac covariant $i\gamma _{5}\gamma \cdot P/M$ besides $\gamma
_{5}$, and get $f_{P}$ as a function of the parameter ratio $B_{1}/B_{0}$.
We then  calibrate $B_{1}/B_{0}$ to reproduce the experimental value of
Kaon decay constant, $f_{K}=159.8MeV$ to get the value $B_{1}/B_{0}=.163.$
Using this value of tuned parameter, we then obtain $%
f_{D}=232.78MeV~(Expt.~222.6\pm 16.7)$, $f_{D_{S}}=295.18MeV(Expt.~294\pm
27) $, which agree with data \cite{22} within the errors. The decay constant
for B-meson predicted in our framework is $f_{B}=191.6MeV$, which is not far
from the recent experimental result for $f_{B}$ \cite{23}. Further our model
predicts $f_{D_{S}}$ value to be around 22{\%} larger than $f_{D}$ value
which is roughly consistent with the prediction of most of the models which
generally predict $f_{D_{S}}$ to be 10{\%} - 25{\%} larger than $f_{D}$ (as
per recent studies in Ref.\cite{23}). These results are demonstrated
 in Table I.

Among all the pseudo-scalar mesons, pion enjoys a special status. The large
difference between the sum of two constituent quark masses and the pion mass
shows that the quark is far off-shell and $q$ could be the same order as the
pion mass. So it seems better to incorporate all the four Dirac covariants.
However, the present experimental condition does not make it economical and
practical to take into account all four Dirac covariants and make a global
fitting, for the experimental precision is in different ranks for the mesons
covered in this paper. On the other hand, from Table I we see that the $%
f_{P} $ value calculated with only the leading order Dirac covariants $%
\gamma _{5}$ and $i\gamma _{5}\gamma \cdot P/M$ rather than all the four
Dirac covariants for pion can give the results different from data by only
about $20\%$. This indicates the moderate contribution from the other two
higher order Dirac covariants for pion. Such a case can be contrasted with
the case of introducing the other LEADING order covariant $i\gamma
_{5}\gamma \cdot P/M$ \ (besides $\gamma _{5}$) for three heavy mesons,
where we see the significant contribution as large as that from $\gamma _{5}$
term since they are equally leading order covariants and the results for $%
f_{P}$ using these leading order covariants for heavier mesons are quite
close to their experimental values. 

These numerical results for $f_{P}$ obtained in our framework with use of
leading order covariants $\gamma _{5}$ and $\gamma _{5}\gamma \cdot P/M$, as
well as those of $f_{V}$ \cite{1} confirm the validity of our power counting
rule, according to which the leading order covariants should contribute
maximum to meson observables, and inspire the possibility to investigate the
higher order terms in order to get better agreement between calculations and
data (when enough precision is obtained), hence help to obtain a better
understanding of the hadron inner structure.

\bigskip
\bigskip

\textbf{\Large Acknowledgements:}

This work has originated from discussions between the authors at ICTP and
their subsequent interactions. It was done within the framework of
Associateship scheme of ICTP. SB wishes to thank Department of Physics, AAU
for providing necessary facilities for this work. SL is supported in part by
National Natural Science Foundation of China (NSFC) with 
grant number 10775090.

\begin{table}[tbp]
\begin{tabular}{|p{3.0cm}|p{1.0cm}|p{1.6cm}|p{1.8cm}|p{2.1cm}|p{1.6cm}|p{1.5cm}|}
\hline
& $B_1/B_0$ & $f_\pi $ \par & $f_K $ \par & $f_D $ & $f_{D_S } $ & $f_B $ \\ 
\hline
BSE-CIA with $\gamma _{5}B_{0}-i\gamma _{5}\gamma \cdot P\frac{B_{1}}{M} $ & 
.17 \par\textbf{.163} \par.16 \par.15 \par.148 \par.14 & 93.0 \par\textbf{%
104.3} \par109.2 \par126.1 \par130.7 \par143.5 & 156.7 \par\textbf{159.8} %
\par160.7 \par164.0 \par165.0 \par168.8 & 229.1 \par\textbf{232.7} \par234.3 %
\par239.7 \par240.0 \par245.3 & 291 \par\textbf{295} \par296 \par303 \par304 %
\par309 & 188 \par\textbf{192} \par193 \par200 \par201 \par206 \\ 
\hline\hline
BSE-CIA with $\gamma_5$ only &  & 138.0 & 153.5 & 137.4 & 159 & 114 \\ 
\cline{1-1}\cline{3-7}
SDE \cite{12} with parameters \par$\omega$, $D$: $45GeV$, $.25GeV^2$ &  &  & 
164 &  &  &  \\ \cline{1-1}\cline{3-7}
SDE \cite{4,11} with \par$\omega (GeV)$, $D(GeV^2)$: $.3$, $1.25$ \par$.4$, $%
.93$ \par$.5$, $.79$ &  &  & \par~ \par~ \par~ \par154 \par155 \par157 &  & 
&  \\ \cline{1-1}\cline{3-7}
Lattice \cite{19} &  &  &  & $208\pm 4$ & 241$\pm 3$ &  \\ 
\cline{1-1}\cline{3-7}
QCD Sum Rule \cite{20} &  &  &  & $203\pm 20$ & $233\pm 23$ &  \\ 
\cline{1-1}\cline{3-7}\cline{1-1}\cline{3-7}
Expt. Results \cite{22} &  & $130.7\pm .1$ & $159.8\pm 1.4$ & $222.6\pm 16.7$
& 294$\pm 27 $ &  \\ \cline{1-1}\cline{3-7}
Babar and Belle Collaboration \cite{23} &  &  &  &  &  & $237\pm 37$ \\ 
\hline
\end{tabular}
\caption{ Leptonic decay constants (in $MeV$) $f_{P}$ in BSE under CIA for
range of values of ratio $B_{1}/B_{0}$. The decay constants are calculated
from data \protect\cite{22}. The masses of hadrons are also from 
\protect\cite{22}. The values of constituent quark masses used are set to be 
$m_{u,d}=300MeV$, $m_{s}=540MeV$, $m_{c}=1500MeV,m_{b}=4500MeV$, which is
compatible with the other parameters fixed from hadron spectrum. Comparisons
with results obtained from other models are also provided. 
As discussed in the paper, we calibrate $B_{1}/B_{0}$ to Kaon data (the 2nd
line), however, if we calibrate $B_{1}/B_{0}$  to pion data
 the results (the 5th line) are as good for the heavier mesons. }
\label{tab1}
\end{table}

\begin{figure}[tbp]
\epsfig{file=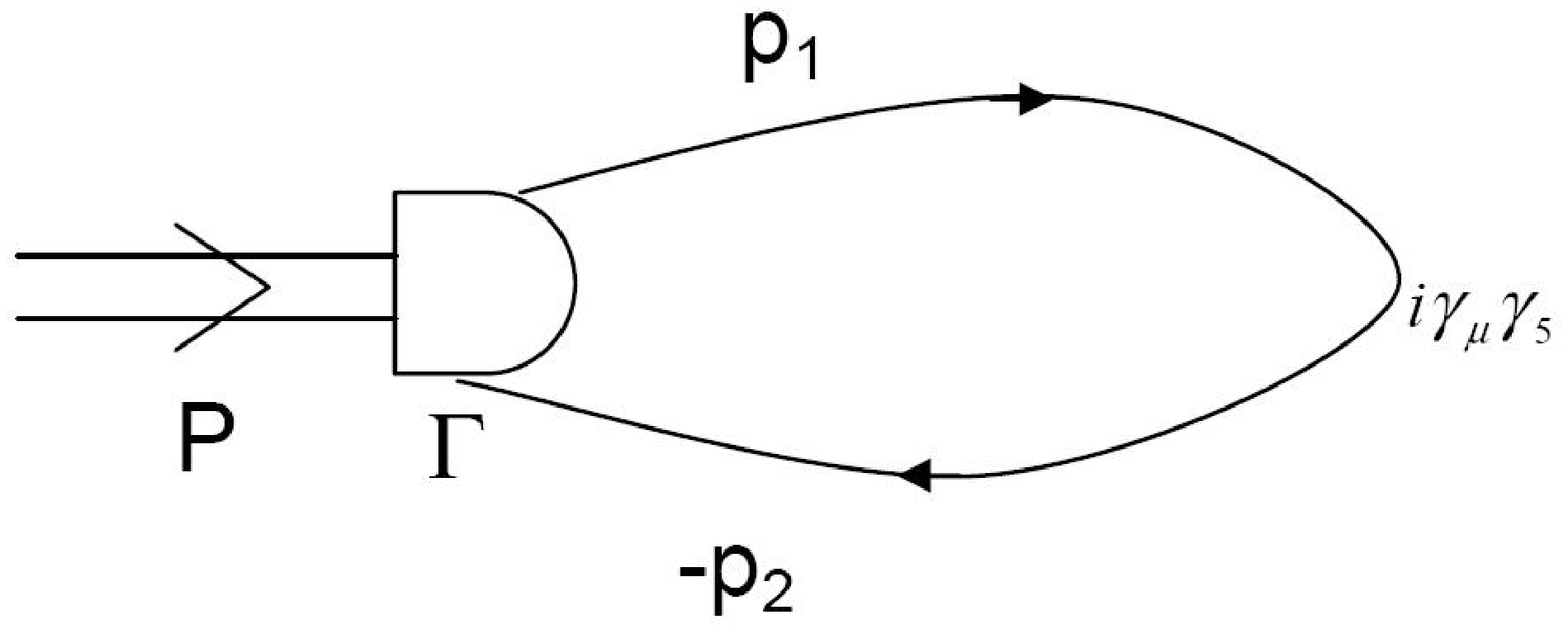,height=8cm,width=13cm}
\caption{}
\label{fig1}
\end{figure}

\end{document}